\documentclass[aoas,MSNbibl,nameyear,dvips]{arximspdf}
\usepackage{mathbh,dcolumn}
\usepackage{stfloats}
\usepackage{graphicx}

\doi{10.1214/13-AOAS701} 
\volume{8}
\issue{1}
\pubyear{2014}
\firstpage{530}
\lastpage{552}

\makeatletter
\fnbelowfloat
\newcolumntype{d}[1]{D{.}{.}{#1}}
\newcommand{\rrvert}{\vert}
\newcommand{\llvert}{\vert}
\newtheorem{teo}{Theorem}
\newproclaim{defn}{Definition}

\newtheorem{algorithm}{Algorithm}
\newproclaim{remark}{Remark}
\newcommand{\SE}{\mathrm{SE}}
\makeatother

\begin{document}
\begin{frontmatter}

\title{Statistical analysis of trajectories on Riemannian manifolds:
Bird migration, hurricane tracking and~video surveillance}
\runtitle{Statistical analysis of trajectories}

\begin{aug}
\author[A]{\fnms{Jingyong} \snm{Su}\corref{}\thanksref{m1}\ead[label=e1]{jingyong.su@ttu.edu}},
\author[B]{\fnms{Sebastian} \snm{Kurtek}\thanksref{m2}},
\author[C]{\fnms{Eric} \snm{Klassen}\thanksref{m3}}\\
\and
\author[C]{\fnms{Anuj} \snm{Srivastava}\thanksref{m3}}
\runauthor{Su, Kurtek, Klassen and Srivastava}
\affiliation{Texas Tech University\thanksmark{m1}, Ohio State
University\thanksmark{m2} and Florida State University\thanksmark{m3}}
\address[A]{J. Su\\
Department of Mathematics and Statistics\\
Texas Tech University\\
Lubbock, Texas 79409\\
USA\\
\printead{e1}} 
\address[B]{S. Kurtek\\
Department of Statistics\\
Ohio State University\\
Columbus, Ohio 43210\\
USA}
\address[C]{E. Klassen\\
A. Srivastava\\
Department of Mathematics\\
Florida State University\\
Tallahassee, Florida 32306\\
USA}
\end{aug}

\received{\smonth{1} \syear{2013}}
\revised{\smonth{11} \syear{2013}}

%
\begin{abstract}
We consider the statistical analysis of trajectories on
Riemannian manifolds that are observed under arbitrary temporal evolutions.
Past methods rely on cross-sectional analysis, with the given temporal
registration,
and consequently may lose the mean structure and artificially inflate
observed variances.
We introduce a quantity that provides both
a cost function for temporal registration and a proper distance for
comparison of trajectories.
This distance is used to define statistical summaries,
such as sample means and covariances, of synchronized trajectories
and ``Gaussian-type'' models to capture their variability at discrete times.
It is invariant to identical time-warpings (or temporal
reparameterizations) of trajectories.
This is based on a novel mathematical representation of trajectories,
termed transported square-root
vector field (TSRVF), and the $\mathbb{L}^2$ norm on the space of TSRVFs.
We illustrate this framework using three representative manifolds---$\mathbb{S}
^2$, $\SE(2)$
and shape space of planar contours---involving both simulated and real
data. In particular,
we demonstrate: (1) improvements in mean structures and significant
reductions in
cross-sectional variances using
real data sets, (2) statistical modeling for capturing variability in
aligned trajectories, and (3) evaluating random
trajectories under these models. Experimental results concern bird
migration, hurricane tracking and
video surveillance.
\end{abstract}

%

\end{frontmatter}

\section{Introduction}\label{sec1} 
\begin{figure}

\includegraphics{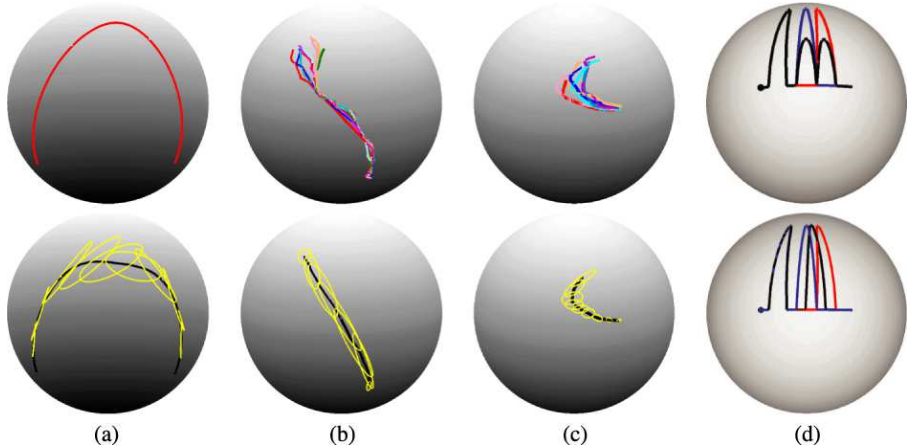}

\caption{Summary of trajectories on $\mathbb{S}^2$: \textup{(a)} a simulated example;
\textup{(b)} bird migration paths; \textup{(c)} hurricane tracks; \textup{(d)} cross-sectional
mean of two trajectories without (top) and with (bottom) registration.}\label{applicationmot}
\end{figure}

The need to summarize and model trajectories
arises in many statistical procedures.
An important issue in this context is that trajectories are often
observed at random times.
If this temporal variability is not accounted for in
the analysis, then the resulting statistical summaries will not be
precise. The mean trajectory may
not be representative of individual trajectories and the
cross-sectional variance will be artificially inflated. This, in turn,
will greatly reduce the effectiveness of any subsequent modeling or
analysis based on the estimated mean and covariance.
As a simple example consider the trajectory on $\mathbb{S}^2$ shown in
the top
panel of Figure~\ref{applicationmot}(a).
We simulate a set of random, discrete observation times and generate
observations
of this trajectory at these random times. These simulated trajectories
are identical in terms of the points traversed but
their evolutions, or parameterizations, are quite different. If we
compute the cross-sectional mean and variance,
the results are shown in the bottom panel. We draw the sample mean
trajectory in black and
the sample variance at discrete times using tangential ellipses. Not
only is the mean fairly different from
the original curve, the variance is purely due to randomness in
observation times and is somewhat artificial.
If we have observed the trajectory at fixed, synchronized times, this
problem would not exist.
%
%

To motivate further, consider the phenomenon of bird migration which
is the regular seasonal journey undertaken by many species of birds.
There are variabilities in migration trajectories, even within the same
species, including the variability
in their rates of travels. In other words, either birds can travel
along different paths or, even if they travel the
same path, different birds (or subgroups) may fly at different speed
patterns along the path. This results in
variability in observation times of migration paths for different birds
and artificially
inflates the cross-sectional variance in the data. Another issue is that
such trajectories are naturally studied as paths on a unit sphere which
is a nonlinear manifold.
We will study the migration data for Swainson's Hawk, with some example
paths shown
in the top panel of Figure~\ref{applicationmot}(b).
Swainson's Hawk inhabits North America mainly in the spring and summer,
and winters in South America.
It shows perhaps the longest migration of any North American raptor,
with durations in excess of two months.
\citet{Owen08} discovered that Swainson's Hawk in migratory
disposition exhibits reduced immune system functions.
Therefore, it becomes important to investigate and summarize such
travels. The bottom panel in Figure~\ref{applicationmot}(b)
shows the cross-sectional sample mean and variance of the trajectories.

Another motivating application comes from hurricane tracking, where one
is interested in studying the shapes of
hurricane tracks in certain geographical regions.
The statistical summaries and models of hurricane tracks can prove very
useful for monitoring and issuing warnings.
Hurricanes potentially evolve at variable dynamical rates and any statistical
analysis to these tracks should be
invariant of the evolution rates. As in the previous application, the
hurricane tracks also are naturally treated
as trajectories on a unit sphere. The top panel of Figure~\ref{applicationmot}(c) shows a set of hurricane tracks originating from
the Atlantic region.
The sample mean and variance of these trajectories are adversely
affected by phase variability,
as shown in the bottom row of Figure~\ref{applicationmot}(c).

As the last motivating example, consider two synthetic trajectories, drawn in red
and blue in the top
of Figure~\ref{applicationmot}(d).
These two trajectories have the same shape, that is,
two bumps each, and a curve representing their
mean is also expected to have two bumps. A simple cross-sectional mean, shown
by the black trajectory in the same picture, has three bumps. If we
solve for the optimal
temporal alignment, then such inconsistencies are avoided and the black
trajectory in
the bottom panel shows the mean obtained using the method proposed
in this paper, which accounts for the time-warping variability.

Although there has been progress in the removal of temporal variability,
often termed \textit{phase variability}, in Euclidean spaces including
\citet{Trouve2000}, \citet{kneip-ramsay2008},
\citet{srivastava-etal-arXiv2011} and \citet
{tucker-wu-srivastava-CSDA2013},
there has not been any treatment of trajectories on Riemannian manifolds.
There are many other applications involving analysis of trajectories on
Riemannian manifolds. For example, human activity recognition has
attracted tremendous interest in
recent years because of its potential in applications such as
surveillance, security and human body animation. There are several
survey articles, for example, \citet{Aggarwal99} and \citet
{Gavrila99}, that provide a detailed
review of research in this area. Here each observed activity is
represented by a sequence of silhouettes in video frames, each silhouette
being an element of the shape space of planar contours. The shape
sequences have
also been called \textit{shape curves} or curves on shape spaces
[\citet{kenobi2010,huiling2003}]. Since activities can be
performed at
different execution rates, their corresponding shape curves will
exhibit distinct evolution rates. \citet{ashok-srivastava-etal-TIP09} accounted for the time-warping
variability but their method has some fundamental problems, as
explained later.
[Briefly, the method is based on equation (\ref{eqregister-past}) which
is not a proper distance.
In fact, it is not even symmetric.] Another motivating
application is in pattern analysis of vehicle trajectories at a
traffic intersection using surveillance videos, where the
instantaneous motion of a vehicle is denoted by the position and
orientation on the road. The movements of vehicles
typically fall into predictable categories---left turn, right turn,
$U$~turn, straight line---but the instantaneous speeds can vary
depending on the traffic. In order to classify these
movements, one has to temporally align the trajectories,
thus removing the effects of travel speeds, and then compare them.

Now we describe the problem in mathematical terms.
Let $\alpha\dvtx  [0,1] \to M$, where
$M$ is a Riemannian manifold, be a differentiable map; it denotes a
trajectory on~$M$. We will
study such trajectories as elements of
an appropriate subset of $M^{[0,1]}$.
Rather than observing
a trajectory $\alpha$ directly, say, in the form of time observations
$\alpha(t_1)$, $\alpha(t_2),\ldots,$ we instead observe the time-warped trajectory
$\alpha(\gamma(t_1))$, $\alpha(\gamma(t_2)),\ldots,$ where
$\gamma\dvtx  [0,1] \to[0,1]$ is an unknown time-warping function
(a~function with certain constraints described later) that governs the
rate of evolution.
The mean and variance of $\{ \alpha_1(t), \alpha_2(t), \ldots, \alpha
_n(t)\}$ for any $t$, where $n$ is the number of observed
trajectories, are termed the cross-sectional mean and variance at that $t$.
If we use the observed samples $\{ \alpha_i(\gamma_i(t)),i=1,2,\ldots,n\}
$ for analysis,
the cross-sectional variance is inflated due to random $\gamma_i$. Our
hypothesis is that this problem can be mitigated
by temporally registering the trajectories.
Thus, we are interested in the following four tasks:
\begin{enumerate}[4.]
\item \textit{Temporal registration}: This is a process of
establishing a one-to-one correspondence between points along multiple
trajectories.
That is, given any $n$ trajectories, say,
$\alpha_1, \alpha_2,\ldots,\alpha_n$,
we are interested in finding functions $\gamma_1, \gamma_2,\ldots,
\gamma_n$
such that the points $\alpha_i(\gamma_i(t))$ are matched optimally for
all $t$.
\item \textit{Metric-based comparisons}:
We want to develop a metric that is invariant to different evolution
rates of trajectories.
Specifically, we want to
define a distance $d(\cdot, \cdot)$ such that for arbitrary evolution
functions
$\gamma_1, \gamma_2$ and arbitrary \mbox{trajectories} $\alpha_1$~and~$\alpha_2$,
we have $d(\alpha_1, \alpha_2) = d(\alpha_1 \circ\gamma_1, \alpha_2
\circ\gamma_2)$.

\item \textit{Statistical summary}: The main use of this metric will be in
defining and computing
a (Karcher) mean trajectory $\mu(t)$ and a cross-sectional variance
function~$\hat\rho(t)$,
associated with any given set of trajectories.
The reason for \mbox{performing} registration is to reduce the
cross-sectional variance that is artificially
introduced in the data due to random observation times. The reduction
in variance is quantified using $\hat\rho$.

\item \textit{Statistical modeling and evaluation}: We will use the
estimated mean and covariance of
registered trajectories to define a ``Gaussian-type'' model on random
trajectories. This model
will then be used to evaluate $p$-values associated with new
trajectories. Here the $p$-value implies the
proportion of trajectories with smaller density than the current trajectory
under the given model.
\end{enumerate}

For performing comparison and summarization of trajectories, we
need a metric and, at first, we consider a more conventional solution.
Since $M$ is a Riemannian manifold, we have a natural distance $d_m$ between
points on $M$. Using $d_m$, one can compare any two trajectories:
$\alpha_1, \alpha_2\dvtx  [0,1] \to M$, as
%
\begin{equation}
d_x(\alpha_1, \alpha_2) = \int
_0^1 d_m\bigl(\alpha_1(t),
\alpha_2(t)\bigr) \,dt. \label{eqsum-metric}
\end{equation}
Although this quantity represents a natural extension of $d_m$ from $M$
to $M^{[0,1]}$, it
suffers from the problem that $d_x(\alpha_1, \alpha_2) \neq
d_x(\alpha_1 \circ\gamma_1, \alpha_2 \circ\gamma_2)$ in general.
It~is not preserved even when the same $\gamma$ is applied to both
trajectories,
that is, $d_x(\alpha_1, \alpha_2) \neq d_x(\alpha_1
\circ\gamma, \alpha_2 \circ\gamma)$ in general. If we have
equality in the last case, for all $\gamma$, then we can develop
a fully invariant distance and use it to properly register
trajectories, as described later. So, the failure to
have this equality is a key issue that forces us to look for
other solutions in situations where trajectories are observed at
random temporal evolutions. When a trajectory $\alpha$ is observed
as $\alpha\circ\gamma$, for an arbitrary temporal
re-parameterization $\gamma$, we call this perturbation \textit{compositional noise}. In these terms, $d_x$ is not useful in
comparing trajectories observed under compositional noise.

Our goal is to take time-warping into account, derive a
warping-invariant metric,
and generate statistical summaries (sample mean, covariance, etc.) for
trajectories on a set $M$.
The fact that $M$ is a Riemannian manifold presents
a formidable challenge in developing a comprehensive framework. But
this is not
the only challenge. To clarify, how has this registration and analysis
problem been handled for trajectories in Euclidean spaces?
In case $M = \mathbb{R}$, that is, if one is interested in
registration and
modeling of real-valued functions under random time-warpings,
the problem has been studied by many authors,
including \citet{srivastava-etal-arXiv2011}, \citet
{mueller-JASA2004}, \citet{kneip-ramsay2008} and \citet
{tucker-wu-srivastava-CSDA2013}.
In case $M=\mathbb{R}^2$, where the problem involves registration and shape
analysis of planar curves,
the solution is discussed in \citet{Michor2007}, \citet
{younes-michor-mumford-shah08}, \citet{Shah2008} and \citet
{sunder-yezzi-siam2011}.
\citet{srivastavaetalPAMI11} proposed a solution that applies to
curves in arbitrary $\mathbb{R}^n$.
One can also draw solutions from problems in image registration where
2D and 3D images are
registered to each other using a spatial warping instead of a temporal
warping [see, e.g., LDDMM
technique, \citet{faisal-LDDDMM05}]. A majority of the existing
methods in Euclidean spaces formulate an
objective function of the type
\[
\min_{\gamma} \biggl( \int_0^1
\bigl| \alpha_1(t) - \alpha_2\bigl(\gamma (t)
\bigr)\bigr|^2 \,dt + \lambda{\mathcal R}(\gamma) \biggr),
\]
where $| \cdot|$ is the Euclidean norm, ${\mathcal R}$ is a regularization
term on the warping function~$\gamma$, and
$\lambda> 0$ is a constant. In the case of a Riemannian manifold, one
can modify the first term to obtain
%
\begin{equation}
\min_{\gamma} \biggl( \int_0^1
\,d_m\bigl(\alpha_1(t), \alpha_2\bigl(\gamma(t)
\bigr)\bigr)^2 \,dt + \lambda{\mathcal R}(\gamma) \biggr),
\label{eqregister-past}
\end{equation}
where $d_m(\cdot, \cdot)$ is the geodesic distance on the manifold.
The main problem with this procedure is that (a) it is not
symmetric, that is, the registration of $\alpha_1$ to $\alpha_2$ is not
the same as that of $\alpha_2$ to $\alpha_1$, as pointed out by
\citet{gary-johnson-MI2001}, among others, and (b) the minimum
value is
not a proper distance, so it cannot be used to compare trajectories.
This sums up the fundamental dilemma in trajectory analysis---equation
(\ref{eqsum-metric}) provides a metric between trajectories but does
not perform registration, while equation (\ref{eqregister-past})
performs
registration but is not a metric.

Another potential approach is to map trajectories onto a vector
space, for example, the tangent space at a point, using the inverse exponential
map, and then compare the mapped trajectories using the Euclidean
solutions in the vector space. While this idea is feasible, the results
may not be consistent since the inverse exponential map is a local
and highly nonlinear operator. For example, on a sphere, under stereographic map two points
near the
south pole will map to two distant points in the tangent space at the
north pole, and their distance will be highly distorted. In contrast, the
solution proposed here transports vector fields associated with
trajectories, rather than trajectories themselves, into a standard tangent
space and this provides a more stable alternative.

We would like an objective function for alignment
that (a) is a proper distance, that is, it is symmetric, positive
definite and satisfies the triangle inequality, (b) is invariant
to simultaneous warping of two trajectories by the same warping
function, and (c) leads to
minimal cross-sectional variance for sample trajectories. For
real-valued functions, a Riemannian framework
has already been presented in \citet{KurtekNIPS} and \citet
{srivastava-etal-arXiv2011}, but to our knowledge this
framework has not been generalized to manifolds.

In this paper we develop a framework for
automated registration of multiple trajectories and obtain improvements in
statistical summaries of time-warped trajectories on Riemannian manifolds.
This framework is based on a novel mathematical representation called
the transported square-root
vector field (TSRVF) and the $\mathbb{L}^2$ norm between TSRVFs.
The setup satisfies the invariance property mentioned earlier, that is,
an identical time-warping of TSRVFs representing two trajectories preserves
the $\mathbb{L}^2$ norm of their difference and, therefore, this
difference is
used to define a warping-invariant
distance between trajectories. The resulting distance is found useful
in registration, comparison and summarization
of trajectories on manifolds. To illustrate these ideas,
we take three manifolds, $\mathbb{S}^2$, $\SE(2)$ and the shape space of
planar closed curves, and provide simulated and real examples.
Our paper can also be viewed as an extension, albeit not a trivial one,
of the work of \citet{KurtekNIPS} and \citet
{srivastava-etal-arXiv2011}
from $M = \mathbb{R}$ to Riemannian manifolds.

The paper is organized as follows. In Section~\ref{sec2} we introduce a general
mathematical framework
for analyzing trajectories on Riemannian manifolds and demonstrate the
use of this
framework in registration, comparison, summarization, modeling and evaluation.
We also provide algorithms for performing these tasks.
In Section~\ref{sec3} we specialize this framework to $\mathbb{S}^2$ and
consider two
applications.
In Section~\ref{sec4} we apply it to pattern analysis of vehicle trajectories on $\SE(2)$.
In Section~\ref{sec5} we provide details for time-warping
invariant analysis of trajectories on the shape space of planar closed
curves, with
applications to activity recognition.

\section{Mathematical framework}\label{sec2}
Let $\alpha$ denote a smooth trajectory on a Riemannian manifold $M$
endowed with a Riemannian metric $\langle\cdot,\cdot\rangle$.
Let ${\mathcal M}$ denote the set of all such trajectories: ${\mathcal
M} =
\{ \alpha\dvtx [0,1] \to M | \alpha$ is smooth$\}$. Also, define
$\Gamma$ to be the set of all orientation preserving diffeomorphisms
of $[0,1]\dvtx  \Gamma=\{\gamma\dvtx [0,1]\to[0,1]|\gamma(0)=0, \gamma(1)=1, \gamma $ is a diffeomorphism$\}$.
Note that $\Gamma$ forms a group under the composition operation. If
$\alpha$ is a trajectory on $M$, then $\alpha\circ\gamma$ is a
trajectory that follows the same sequence of points as $\alpha$ but
at the evolution rate governed by~$\gamma$. More technically, the
group $\Gamma$ acts on ${\mathcal M}$ according to $(\alpha,\gamma
)=\alpha
\circ\gamma$.

Given two smooth trajectories $\alpha_1,  \alpha_2 \in{\mathcal
M}$, we
want to
register points along the trajectories and compute a time-warping
invariant distance between them. As mentioned earlier, the quantity
given in equation (\ref{eqregister-past}) would be a natural choice for
this purpose, but
it fails for several reasons, including the fact that it is not symmetric.
Fundamentally, this and other quantities used in previous literature
are not appropriate
for solving the registration problem because they are not measuring registration
in the first place. To highlight this issue, take the registration of
points between the
pair $(\alpha_1, \alpha_2)$ and the pair $(\alpha_1 \circ\gamma,
\alpha
_2 \circ\gamma)$,
for any $\gamma\in\Gamma$.
It can be seen that
the pairs $(\alpha_1, \alpha_2)$ and
$(\alpha_1 \circ\gamma, \alpha_2 \circ\gamma)$ have exactly the same
registration of points.
In fact,
any identical time-warping of two trajectories does not change
the registration of points between them.
But the quantities given in
equations (\ref{eqsum-metric}) and (\ref{eqregister-past}) provide
different values for these pairs, despite
the same registration. Hence,
they are not good measures of registration.
We emphasize that the invariance under identical time-warping is a key
property needed in the desired
framework.

We introduce a new representation of trajectories that will be used to
compare and register them.
We assume that for any two points $p, q \in M$, we have an expression
for parallel transporting any vector
$v \in T_p(M)$ along the shortest geodesic from $p$ to~$q$, denoted by $(v)_{p
\rightarrow q}$.
As long as $p$ and $q$ do not fall in the cut loci of each other, the
geodesic between them is unique and
the parallel transport is well defined. The measure of the set of cut
locus on the manifolds of our interest is typically zero.
So, the practical implications of this limitation are negligible.
Let $c$ be a point in~$M$ that we designate as a reference point.
We assume that none of the observed trajectories pass through the cut
locus of $c$ to avoid the problem mentioned above.
%
%
\begin{defn}\label{defn1}
For any smooth trajectory $\alpha\in{\mathcal M}$, the transported
square-root vector field (TSRVF) is
a parallel transport of a scaled velocity vector field of $\alpha$ to a
reference point $c \in M$ according to
\[
h_{\alpha}(t) = \frac{ \dot{\alpha}(t)_{\alpha(t) \rightarrow c} }{
\sqrt { | \dot{\alpha}(t) |} } \in T_{c}(M),
\]
where
$| \cdot|$ denotes the norm related to the Riemannian metric on $M$.
\end{defn}
Since $\alpha$ is smooth, so is the vector field $h_{\alpha}$. Let
${\mathcal H} \subset T_c(M)^{[0,1]}$ be the set of smooth curves in
$T_c(M)$ obtained as TSRVFs of trajectories in $M$, ${\mathcal H} = \{
h_{\alpha} | \alpha\in{\mathcal M}\}$.
If $M = \mathbb{R}^n$ with the Euclidean metric, then $h$ is exactly the
square-root velocity function defined in \citet{srivastavaetalPAMI11}.

The choice of the reference point $c$ used in
Definition \ref{defn1} is important and can affect the
results. The choice typically depends on the application, the data and
the manifold under study.
In case all the trajectories pass through a point or pass close to a
point, then that point
is a natural candidate for $c$. This would be true, for example, in the
case of hurricane tracks, if we
are focused on all hurricanes starting from the same region.
Another remark is that instead of parallel transporting of scaled
velocity vectors along geodesics,
one can transport them along trajectories themselves, as was done by
\citet{Jupp1987},
but that requires $c$ to be a common point of all trajectories.
While the choice of $c$ can, in principle, affect distances,
our experiments suggest that the results of registration,
distance-based clustering and classification
are quite stable with respect to this choice. An example is presented
later in Figure~\ref{fig1Align-1}.

We represent a trajectory $\alpha\in{\mathcal M}$ with the pair
$(\alpha
(0), h_{\alpha})
\in M \times{\mathcal H}$. Given this representation, we can reconstruct
the path,
an element of ${\mathcal M}$, as follows. For any time $t$, let $V_t$
be a
time-varying
tangent vector-field on $M$ obtained by parallel transporting
$h_{\alpha
}(t)$ over $M$
[except for the cut locus of $\alpha(t)$], that is, for any $p \in M$,
$V_p(t) = (h_{\alpha}(t))_{c \rightarrow p}$.
Then, define an integral curve $\beta$ such that $\dot{\beta}(t) =
|V_{\beta(t)}(t)|V_{\beta(t)}(t)$ with the starting point $\beta(0) = \alpha(0) \in M$.
This resulting curve $\beta$ will be exactly the same as the original
curve $\alpha$.

The starting points of different curves can be compared using
the Riemannian distance $d_m$ on $M$. However, these points do not play
an important
role in the alignment of trajectories since they are already assumed to
be matched to
each other. Therefore, the main focus of analysis, in terms of
alignment and comparison, is on TSRVFs. Since a TSRVF is a path in
$T_c(M)$, one can use the
$\mathbb{L}^2$ norm to compare such paths.
%
%
\begin{defn} \label{defn2}
Let $\alpha_1$ and $\alpha_2$ be two smooth trajectories on $M$ and let
$h_{\alpha_1}$ and $h_{\alpha_2}$ be the corresponding TSRVFs. The
distance between them is
\[
d_h(h_{\alpha_1}, h_{\alpha_2}) = \biggl( \int
_0^1 \bigl| h_{\alpha
_1}(t) - h_{\alpha_2}(t)
\bigr|^2 \,dt \biggr)^{1/2}.
\]
\end{defn}
The distance $d_h$, being the standard $\mathbb{L}^2$ norm, satisfies
symmetry, positive definiteness and triangle inequality.
Also, due to the invertibility of the mapping from~${\mathcal M}$
to $M \times{\mathcal H}$, one can use $d_h$ (along with $d_m$) to define
a distance on ${\mathcal M}$.
The main motivation of this setup---TSRVF representation and $\mathbb{L}^2$
norm---comes from the
following fact.
If a trajectory $\alpha$ is warped by $\gamma$, to result in $\alpha
\circ\gamma$,
the TSRVF of $\alpha\circ\gamma$ is given by
\begin{eqnarray*}
h_{\alpha\circ\gamma}(t) &=& \frac{ (\dot{\alpha}(\gamma(t)) \dot{\gamma
}(t))_{\alpha(\gamma(t)) \rightarrow c} }{\sqrt{ | \dot{\alpha
}(\gamma(t)) \dot{\gamma}(t)|} } = \frac{ (\dot{\alpha}(\gamma(t)))_{\alpha(\gamma(t)) \rightarrow c}
\sqrt{\dot
{\gamma}(t)} }{\sqrt{ | \dot{\alpha}(\gamma(t)) |} }
\\
&=& h_\alpha
\bigl(\gamma(t)\bigr) \sqrt{\dot{\gamma}(t)},
\end{eqnarray*}
which is also denoted as $(h_\alpha, \gamma)(t)$. We will often write
$(h_{\alpha},\gamma)$ to denote $h_{\alpha\circ\gamma}$.
As stated earlier, we need a distance for registration that is
invariant to identical time-warpings of trajectories.
Next, we show that $d_h$ satisfies this property.
%
%
\begin{teo} \label{theoisometry}
For any $\alpha_1, \alpha_2 \in{\mathcal M}$ and $\gamma\in\Gamma
$, the distance
$d_h$ satisfies
$d_h(h_{\alpha_1\circ\gamma}, h_{\alpha_2 \circ\gamma)} =
d_h(h_{\alpha_1}, h_{\alpha_2})$.
In geometric terms, this implies that the action of $\Gamma$ on
${\mathcal
H}$ under the
$\mathbb{L}^2$ metric is by isometries.
\end{teo}
The proof is given below:
\begin{eqnarray*}
d_h(h_{\alpha_1\circ\gamma}, h_{\alpha_2 \circ\gamma})&=& \biggl( \int
_0^1 \bigl| h_{\alpha_1}\bigl(\gamma(t)\bigr)
\sqrt{\dot{\gamma}(t)} - h_{\alpha_2}\bigl(\gamma(t)\bigr) \sqrt{\dot{
\gamma}(t)} \bigr|^2 \,dt \biggr)^{1/2}
\\
&=& \biggl( \int_0^1 \bigl| h_{\alpha_1}(s) -
h_{\alpha_2}(s) \bigr|^2 \,ds \biggr)^{1/2} =
d_h(h_{\alpha_1}, h_{\alpha_2}),
\end{eqnarray*}
where $s = \gamma(t)$.

Next we define a quantity that can be used as a distance between
trajectories while being invariant to their temporal variability. To
set up this definition, we first introduce an equivalence
relation between trajectories. For any two trajectories $\alpha_1$~and~$\alpha_2$, we define them to be equivalent, $\alpha_1 \sim
\alpha_2$, when:
\begin{enumerate}[1.]
\item$\alpha_1(0) = \alpha_2(0)$, and
\item there exists a sequence $\{\gamma_k\} \in\Gamma$ such
that $\lim_{k \to\infty} h_{(\alpha_1 \circ\gamma_k)} =
h_{\alpha_2}$ under the $\mathbb{L}^2$ metric.
\end{enumerate}
In other words, any two trajectories are equivalent if they have the
same starting
point and the TSRVF of one can be
time-warped into the TSRVF of the other using a sequence of
warpings.
It can be easily checked that $\sim$ forms an equivalence relation on
${\mathcal H}$
(and, correspondingly, ${\mathcal M}$).

Since we want our distance to be invariant to time-warpings of trajectories,
we wish to compare trajectories by comparing their equivalence classes. Thus,
our next step is to inherit the distance $d_h$ to the set of such
equivalence classes.
Toward this goal, we introduce the set $\tilde{\Gamma}$ as the set of
all nondecreasing,
absolutely continuous functions $\gamma\dvtx [0,1] \to[0,1]$ such that
$\gamma(0) = 0$
and $\gamma(1)=1$. This set $\tilde{\Gamma}$
is a semigroup with the composition operation (it is not a group
because the elements do not
have inverses). The group $\Gamma$ is a subset of $\tilde{\Gamma}$. The
elements of $\tilde{\Gamma}$
warp the time axis of trajectories in $M$ in the same way as elements
of $\Gamma$, except they
allow certain singularities.
For a TSRVF $h_{\alpha} \in{\mathcal H}$, its equivalence class, or
\textit{orbit} under $\tilde{\Gamma}$, is given by
$[h_{\alpha}]=\{(h_{\alpha},\gamma) | h_{\alpha}\in{\mathcal H}, \gamma\in
\tilde{\Gamma}\}$.

It can be shown that the orbits under $\tilde{\Gamma}$ are exactly the
same as the closures of the
orbits of $\Gamma$, defined as $[h_{\alpha}]_0 = \{(h_{\alpha
},\gamma)
| \gamma\in\Gamma\}$,
as long as $\alpha$ has nonvanishing derivatives almost everywhere.
(The last
condition is not restrictive since we can always re-parameterize
$\alpha
$ by the arc-length.)
The closure is
with respect to the $\mathbb{L}^2$ metric on ${\mathcal H}$.
Please refer to \citet{dan-robinson-thesis} for a detailed
description of a similar construction for
trajectories in $\mathbb{R}$.

Now we define the quantity that will serve both as the cost function for
registration and distance for comparison.
This quantity is essentially $d_h$ measured between equivalence classes.
%
%
\begin{defn}
The distance $d_s$ on ${\mathcal H}/\sim$ (or ${\mathcal M}/\sim$) is the
shortest $d_h$ distance
between equivalence classes in ${\mathcal H}$, given as
%
\begin{eqnarray}\label{eqndist}
&& d_s\bigl([h_{\alpha_1}], [h_{\alpha_2}]\bigr)\nonumber
\\
&&\qquad =  \inf _{\gamma_1, \gamma_2\in
\tilde{\Gamma}} \,d_h\bigl((h_{\alpha_1},
\gamma_1), (h_{\alpha_2}, \gamma_2) \bigr)
\\
&&\qquad = \inf_{\gamma_1, \gamma_2 \in\tilde{\Gamma}} \biggl( \int_0^1
\bigl| h_{\alpha_1}\bigl(\gamma_1(t)\bigr)\sqrt{\dot{
\gamma_1}(t)} - h_{\alpha_2}\bigl(\gamma_2(t)\bigr)
\sqrt{\dot{\gamma_2}(t)} \bigr|^2 \,dt \biggr)^{1/2}.\nonumber
\end{eqnarray}
\end{defn}

%
\begin{teo}
The distance $d_s$ is a proper distance on ${\mathcal H}/\sim$.
\end{teo}

\begin{pf}
The symmetry of $d_s$ comes directly from the symmetry of $d_h$. For
positive definiteness,
we need to show that $d_s([h_{\alpha_1}],[h_{\alpha_2}])=0\Rightarrow
[h_{\alpha_1}] =[h_{\alpha_2}]$.
Suppose that $d_s([h_{\alpha_1}],[h_{\alpha_2}]) = 0$, by definition,
it then follows immediately that for all
$\varepsilon>0$, there exists a $\gamma\in\Gamma$ such that
$d_h(h_{\alpha_1}, (h_{\alpha_2}, \gamma))
< \varepsilon$. From this, it follows that $h_{\alpha_1}$ is in the orbit
$h_{\alpha_2}$. Since we are assuming that orbits are closed, it
follows that $h_{\alpha_1} \in[h_{\alpha_2}]$, so $[h_{\alpha_1}]=
[h_{\alpha_2}]$.

To establish the triangle inequality,
we need to prove
\[
d_s\bigl([h_{\alpha_1}], [h_{\alpha_3}]\bigr) \leq
d_s\bigl([h_{\alpha_1}], [h_{\alpha_2}]\bigr)+
d_s\bigl([h_{\alpha_2}], [h_{\alpha_3}]\bigr)
\]
for any $h_{\alpha_1}, h_{\alpha_2}, h_{\alpha_3}\in{\mathcal H}$.
For a contradiction, suppose
\[
d_s\bigl([h_{\alpha_1}],[h_{\alpha_3}]\bigr)>
d_s\bigl([h_{\alpha_1}],[h_{\alpha_2}]\bigr)+ \tilde d
\bigl([h_{\alpha_2}],[h_{\alpha_3}]\bigr).
\]
Let
\[
\varepsilon=\tfrac{1}{3}\bigl(d_s\bigl([h_{\alpha_1}],
[h_{\alpha_3}]\bigr)- d_s\bigl([h_{\alpha_1}],[h_{\alpha_2}]
\bigr)- d_s\bigl([h_{\alpha_2}],[h_{\alpha
_3}]\bigr)\bigr).
\]
By our supposition, $\varepsilon>0$. From the definition of $\varepsilon$, it
follows that
\[
d_s\bigl([h_{\alpha_1}],[h_{\alpha_3}]\bigr)=
d_s\bigl([h_{\alpha_1}],[h_{\alpha_2}]\bigr)+
d_s\bigl([h_{\alpha_2}],[h_{\alpha
_3}]\bigr)+3\varepsilon.
\]
By the definition of $d_s$, we can choose $\gamma_1, \gamma_2 \in
\Gamma$,
such that
\[
d_h\bigl((h_{\alpha_1}, \gamma_1),h_{\alpha_2}
\bigr) \leq d_s\bigl([h_{\alpha_1}],[h_{\alpha_2}]\bigr)+
\varepsilon
\]
and
\[
d_h\bigl(h_{\alpha_2},(h_{\alpha_3},\gamma_2)
\bigr) \leq d_s\bigl([h_{\alpha_2}],[h_{\alpha_3}]\bigr)+
\varepsilon.
\]
Now, by the triangle inequality for $d_h$, we know that
\begin{eqnarray*}
d_h\bigl((h_{\alpha_1},\gamma_1),(h_{\alpha_3},
\gamma_2)\bigr) & \leq& d_h\bigl((h_{\alpha_1},
\gamma_1), h_{\alpha_2}\bigr)+d_h
\bigl(h_{\alpha
_2},(h_{\alpha
_3},\gamma_2)\bigr)
\\
& \leq& d_s\bigl([h_{\alpha_1}], [h_{\alpha_2}]\bigr)+
d_s\bigl([h_{\alpha
_2}],[h_{\alpha_3}]\bigr)+2\varepsilon.
\end{eqnarray*}
It follows that
\[
d_s\bigl([h_{\alpha_1}], [h_{\alpha_3}]\bigr) \leq
d_s\bigl([h_{\alpha_1}],[h_{\alpha_2}]\bigr)+
d_s\bigl([h_{\alpha
_2}],[h_{\alpha
_3}]\bigr)+2\varepsilon.
\]
But this contradicts the fact that
\[
d_s\bigl([h_{\alpha_1}],[h_{\alpha_3}]\bigr)=
d_s\bigl([h_{\alpha_1}],[h_{\alpha_1}]\bigr)+
d_s\bigl([h_{\alpha_2}],[h_{\alpha_3}]\bigr)+3\varepsilon.
\]
Hence, our supposition that
\[
d_s\bigl([h_{\alpha_1}],[h_{\alpha_3}]\bigr)>
d_s\bigl([h_{\alpha_1}],[h_{\alpha_2}]\bigr)+
d_s\bigl([h_{\alpha_2}],[h_{\alpha_3}]\bigr)
\]
must be false.
The triangle inequality follows.
\end{pf}

Now, since $\Gamma$ is dense in $\tilde{\Gamma}$, for any $\delta> 0$,
there exists
a $\gamma^*$ such that
%
\begin{equation}
\bigl\llvert d_h( h_{\alpha_1}, h_{\alpha_2 \circ\gamma^*}) -
d_s\bigl({[h_{\alpha_1}]}, {[h_{\alpha_2}]}\bigr)\bigr
\rrvert < \delta. \label{eqndefn-opt-gamma}
\end{equation}
This $\gamma^*$ may not
be unique but any such $\gamma^*$ is sufficient for our purpose.
Furthermore, since $\gamma^* \in\Gamma$, it has
an inverse that can be used in further analysis.
The minimization over $\Gamma$ in equation (\ref{eqndefn-opt-gamma}) is
performed in
practice using the dynamic
programming (DP) algorithm [\citet{Bertsekas2007}]. Here one
samples the interval $[0,1]$
using $T$ discrete points and then restricts to only piecewise linear
$\gamma$'s
that pass through that $T \times T$ grid.
The search for the optimal trajectory on this grid is accomplished in
$O(T^2)$ steps.

\subsection{Metric-based comparison of trajectories}\label{sec2.1}
Our goal of warping-invariant comparison of trajectories is achieved
using $d_s$.
For any $\gamma_1$, $\gamma_2 \in\Gamma$ and $\alpha_1$, \mbox{$\alpha_2
\in {\mathcal M}$,} we have
\[
[h_{\alpha_1 \circ\gamma_1}] = [h_{\alpha_1}],\qquad [h_{\alpha_2
\circ
\gamma_2}] =
[h_{\alpha_2}]
\]
and, therefore, we get $d_s([h_{\alpha_1 \circ\gamma_1}], [h_{\alpha_2
\circ\gamma_2}])
= d_s([h_{\alpha_1}],[h_{\alpha_2}])$.
Examples of this metric are presented later.

\subsection{Pairwise temporal registration of trajectories}\label{sec2.2}
The next goal is to perform registration of points along trajectories.
Let our approximation to the optimal warping be as defined in equation
(\ref{eqndefn-opt-gamma}).
This allows for the registration between $\alpha_1$ and $\alpha_2$,
in that
the point $\alpha_1(t)$ on the first trajectory is optimally matched to
the point
$\alpha_2(\gamma^*(t))$ on the second trajectory.

If we compare equation (\ref{eqndist}) with equation (\ref{eqregister-past}),
we see the advantages of the proposed framework. Both equations present
a registration
problem between $\alpha_1$~and~$\alpha_2$, but only the minimum value
resulting from
equation (\ref{eqndist}) is a proper distance. Also, in equation (\ref{eqregister-past})
we have two separate terms for matching and regularization, with an
arbitrary weight $\lambda$,
but in equation (\ref{eqndist}) the two terms have been merged into a
single natural form.
Recall that the change in TSRVF $h$ due to the time-warping of $\alpha$
by $\gamma$
is given by $(h, \gamma)= (h \circ\gamma)\sqrt{\dot\gamma}$, and the
distance $d_s$ is based on these warped TSRVFs.
The term $\sqrt{\dot\gamma}$ provides an intrinsic regularization on
$\gamma$ in the matching process.
It provides an elastic penalty against excessive warping since $\dot
{\gamma}$ becomes large at those places.
Lastly, the optimal registration
in equation (\ref{eqndist}) remains the same if we change the order of
the input functions. That is, the registration
process is inverse consistent.

\subsection{Summarization and registration of multiple trajectories}\label{sec2.3}
An additional advantage of this framework is that one can compute an
average of several trajectories and use it
as a \textit{template} for future classification. Furthermore, this template
can be used for registering multiple trajectories. We use the notion
of the Karcher mean to define and compute average trajectories.
Given a set of sample trajectories $\alpha_1,\ldots,\alpha_n$ on $M$, we
represent them using the corresponding
pairs $(\alpha_1(0), h_{\alpha_1}), (\alpha_2(0), h_{\alpha_2}),\ldots,
(\alpha_n(0), h_{\alpha_n})$. We
compute the Karcher means of each component in their respective spaces:
(1) the Karcher mean of $\alpha_i(0)$
is computed with respect to $d_m$ in $M$, and (2) the Karcher mean of
$h_{\alpha_i}$ with
respect to $d_s$ in ${\mathcal H}/\sim$. The latter Karcher mean is
defined as
\[
h_{\mu} = \mathop{\operatorname{argmin}}_{[h_{\alpha}] \in{\mathcal H}/\sim
} \sum
_{i=1}^n d_s\bigl([h_{\alpha}],
[h_{\alpha_i}]\bigr)^2.
\]
Note that $[h_{\mu}]$ is an equivalence class of trajectories and one
can select
any element of this mean class to help in the alignment of multiple
trajectories.
The standard algorithm to compute the Karcher mean proposed by
\citet{huiling2000} is adapted to this
problem as follows:
%
%
\begin{algorithm}[(Karcher mean of multiple trajectories)]\label{algo1}
Compute the Karcher mean of $\{\alpha_i(0)\}$ and set it to be $\mu(0)$.
\begin{enumerate}
\item Initialization step: select $\mu$ to be one of the original
trajectories and compute its TSRVF $h_{\mu}$.

\item Align each $h_{\alpha_i}, i=1,\ldots,n$, to $h_{\mu}$ according
to equation (\ref{eqndefn-opt-gamma}). That is, solve for~$\gamma_i^*$ using the DP
algorithm and
set $\tilde{\alpha}_i = \alpha_i \circ\gamma_i^*$.

\item Compute TSRVFs of the warped trajectories, $h_{\tilde{\alpha
}_i}$, $i=1,2,\ldots,n$, and update
$h_{\mu}$ as a curve in $T_c(M)$ according to
$
{h_{\mu}}(t) = \frac{1}{n} \sum_{i=1}^n h_{\tilde{\alpha_i}}(t)$.

\item Define ${\mu}$ to be the integral curve associated with a time-varying
vector field on $M$ generated using $h_{\mu}$, that is, $\frac{d\mu(t)}{dt} =
|{h_{\mu}}(t)| ({h_{\mu}})(t)_{c \rightarrow {\mu}(t)} $,
and the initial condition~$\mu(0)$.

\item
Compute $E=\sum_{i=1}^n {d_s([h_{\mu}], [h_{\alpha_i}])^2} = \sum_{i=1}^n {d_h(h_{\mu}, h_{\tilde{\alpha}_i})^2}$
and check for convergence. If not converged, return to step~2.
\end{enumerate}
\end{algorithm}

It can be shown that the cost function decreases iteratively and, as
zero is a
natural lower bound, $\sum_{i=1}^n d_s([h_{\mu}], [h_{\alpha_i}])^2$
will always converge. This algorithm provides two sets of outputs: an
average trajectory denoted by the final $\mu$ and the set of aligned
trajectories $\tilde{\alpha}_i$. Therefore, this solves the problem
of aligning multiple trajectories too.

For computing and analyzing the second and higher moments of a sample
trajectory,
the tangent space $T_{\mu(t)}(M)$, for $t \in[0, 1]$, is used.
This is convenient because it is a vector space and one
can apply more traditional methods here. First, for each aligned
trajectory $\tilde{\alpha}_i(t)$ at time $t$,
the vector $v_i(t) \in T_{\mu(t)}(M)$ is computed such that a geodesic
that goes from $\mu(t)$ to $\tilde{\alpha}_i(t)$
in unit time has the initial velocity $v_i(t)$. This is also called the
\textit{shooting vector} from $\mu(t)$ to $\tilde{\alpha}_i(t)$.
Let $\hat{K}(t)$ be the sample covariance matrix of all shooting
vectors from $\mu(t)$'s to $\tilde{\alpha}_i(t)$'s.
The sample Karcher covariance\vspace*{-2pt} at time $t$ is given by $\hat{K}(t)=\frac{1
}{n-1}\sum_{i=1}^n v_i(t){v_i(t)}^T$, with the trace
$\hat{\rho}(t) = \operatorname{trace}(\hat{K}(t))$. This $\hat{\rho}(t)$
represents a quantification of the cross-sectional
variance, as a function of $t$, and can be used to study the level of
alignment of trajectories.
Also, for capturing the essential variability in the data, one can
perform Principal Component Analysis (PCA) of the shooting vectors.
The basic idea is to compute the Singular Value Decomposition (SVD)
$\hat{K}(t)=U(t) \Sigma(t) U^T(t)$, where $U(t)$ is an orthogonal
matrix and $\Sigma(t)$
is the diagonal matrix of singular values. Assuming that the entries
along the diagonal in $\Sigma(t)$ are organized
in a nonincreasing order, the functions $U_1(t), U_2(t), \ldots$ represent the
dominant directions of variability in the data.

\subsection{Modeling and evaluation of trajectories}\label{sec2.4}
An important use of means and covariances of trajectories is in
devising probability
models for capturing the observed statistical variability, and for
using these models in
evaluating $p$-values of future observations. By $p$-values we mean the
proportion of random trajectories that will have lower probability
density under a given model
when compared to the test trajectory. Several models are possible in
this situation, but
since our main focus is on temporal registration of trajectories, we
will choose a simple
model to demonstrate our ideas.
After the registration, we treat a trajectory~$\alpha$ as a
discrete-time process, composed of $m$ points as
$\{\alpha(t_1), \alpha(t_2),\ldots,\alpha(t_m)\}$, for a fixed
partition $\{0=t_1, t_2, \ldots, t_m=1\}$ of $[0,1]$.
Given the mean and the covariance at
each $t_j$, we model the points $\alpha(t_j) \in M, j=1, 2,\ldots, m$
independently, and obtain the
joint density by taking the product. The difficulty in this step comes
from the fact that $M$ is a nonlinear manifold but one can use
the tangent space $T_{\mu(t_j)}(M)$, instead, to impose a probability
model since this is a vector space.
We impose a multivariate normal density on the tangent vector $v(t_j) =
\exp_{\mu(t_j)}^{-1}(\alpha(t_j))$,
with mean zero and variance given by $\hat{K}(t_j)$ (as defined above).
It is analogous to the model of additive white Gaussian noise when
$M=\mathbb{R}$.
Then, for any trajectory $\alpha$, one can compute a joint probability\vspace*{1pt}
of the
full trajectory as $P(\alpha)=\prod_{j=1}^m f(\alpha(t_j)) \equiv
\prod_{j=1}^mN(v(t_j); 0, \hat{K}(t_j))$.
This model is potentially useful for many situations: (1) It can be
used to simulate new trajectories via random sampling.
Given $\{(\mu(t_j), \hat{K}(t_j))| t \in[0,1]\}$, we can simulate the
tangent vectors and compute the corresponding trajectory
points $\alpha(t_j)$, for the desired $t_j$.
(2) Given a trajectory, we evaluate its $p$-value under the imposed
model. This
measures how likely is the occurrence of the trajectory by chance
assuming the null hypothesis~$H_0$, where
$H_0$ represents that imposed model.

Since we are interested in studying the effects of temporal
registration, we demonstrate these ideas with the
following experiment. We compute $p$-values of trajectories using the
parametric bootstrap under two situations:
without registration and with registration. In each situation, we first
take a set of trajectories as the training set and
estimate the mean and covariance at each discrete time,
and then impose a ``Gaussian-type'' model on the tangent spaces of $M$
at the mean values at those times.
This becomes the imposed model $H_0$. The evaluation of $p$-values
requires Monte Carlo sampling.
We generate a large number, say, $N={}$10,000, of trajectories from the
model, denoted as $X_i, i=1,2,\ldots,N$.
Then we compute the proportion that are less likely than our test trajectory
and\vspace*{1pt} denote it as $p(\alpha)= \frac{1}{N}\sum_{i=1}^N \mathbh
{1}_{ [ {P(X_i)<P(\alpha)}  ] }$.

In the following sections we consider three examples of
$M$ and present experimental results to validate our framework.

\section{Trajectories on $\mathbb{S}^2$}\label{sec3}
Statistical methods for unit vectors in three-dimensio\-nal space have
been studied extensively in directional statistics [\citet
{mardiadirectional-statistics}].
In the landmark-based shape analysis of objects, including \citet
{dryden-mardiabook98}, \citet{Jupp1987} and \citet{Kume2007},
where 2D objects are represented by configurations of salient points or
landmarks,
the set of all such configurations after removing translation and
scale is a real sphere $\mathbb{S}^{2n-3}$ (for configurations with $n$
landmarks). To illustrate this framework, in a~simple setting, we
start with $M=\mathbb{S}^2$, with the standard Euclidean Riemannian metric.
For any two points $p, q \in\mathbb{S}^2$ ($p \neq-q$) and a
tangent vector
$v \in
T_p(\mathbb{S}^2)$, the parallel transport $(v)_{p \rightarrow q}$
along the
shortest geodesic (i.e., great circle) from $p$ to $q$ is given by
$v-\frac{2 \langle v,q  \rangle}{|p+q|^2}(p+q)$.\vspace*{9pt}

\begin{figure}[t]

\includegraphics{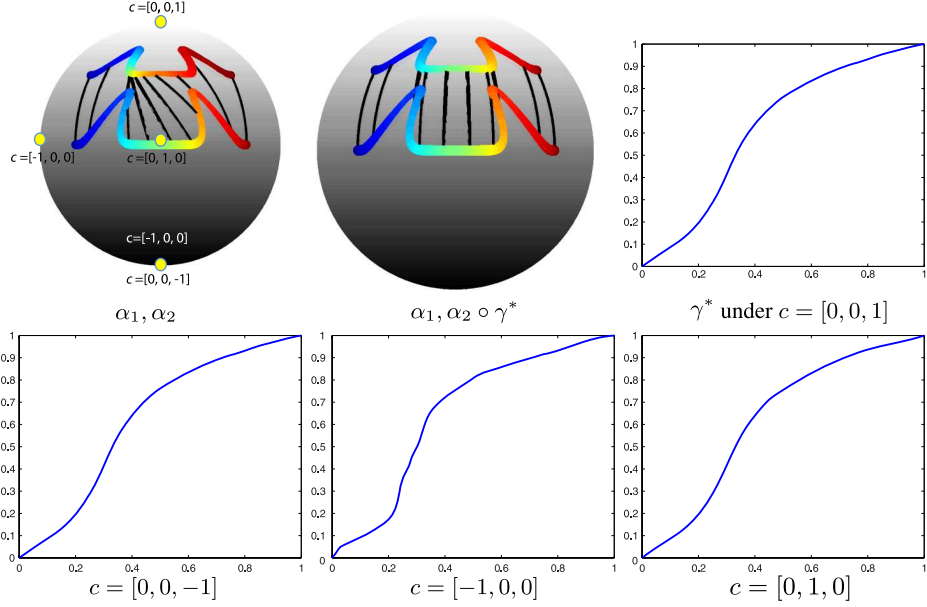}

\caption{Registration of trajectories on $\mathbb{S}^2$.} \label{fig1Align-1}
\end{figure}

\textit{Registration of trajectories}: As mentioned earlier,
for any two trajectories on~$\mathbb{S}^2$, we use their TSRVFs and DP
algorithm in equation (\ref{eqndefn-opt-gamma}) to find the optimal registration
between them. In Figure~\ref{fig1Align-1} we show one example of
registering such trajectories. The parameterization of trajectories
is displayed using colors. In the top row, the left column shows
the trajectories $\alpha_1$ and $\alpha_2$, the middle column
shows $\alpha_1$ and $\alpha_2 \circ\gamma^*$ and the right column
shows $\gamma^*$ using $c=[0, 0, 1]$. The correspondences between the
two trajectories are
depicted by black lines connecting points along them. Due to the
optimization of $\gamma$ in equation (\ref{eqndefn-opt-gamma}), the
$d_h$ value
between them reduces from 1.67 to 0.36 and the correspondences become
more natural after the
alignment. We also consider different choices of $c$ ($c=[0, 0, -1],
[-1, 0, 0], [0, 1, 0]$).
In all cases the registration results are very close, as shown in the bottom row.

%
%
%
\begin{figure}[b]

\includegraphics{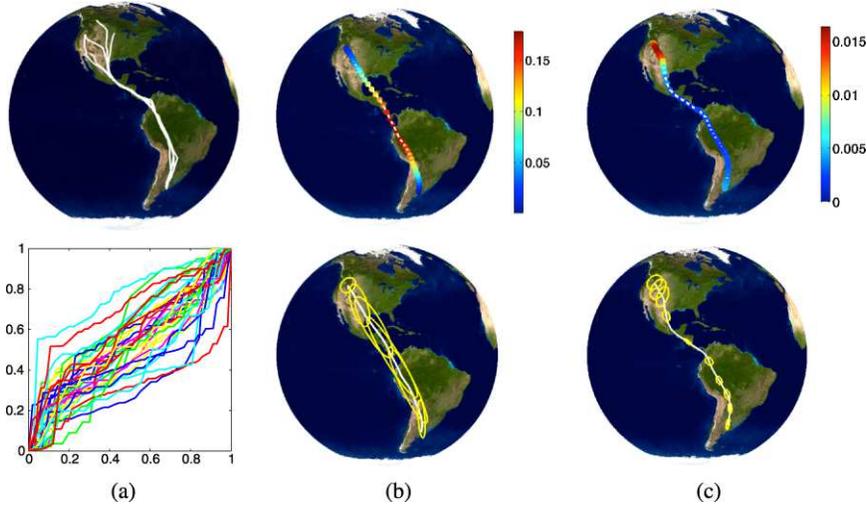}

\caption{Swainson's Hawk migration: \textup{(a)} $\{\alpha_i\}$ (top) and $\{\gamma_i^*\}$ (bottom);
\textup{(b)} $\mu$ and $\hat{\rho}$ without registration; \textup{(c)} $\mu$ and $\hat{\rho}$ with registration.}\label{birdmigration}
\end{figure}

In the following, we consider two specific applications, bird migration
and hurricane tracks, and show how
the cross-sectional variance of the mean trajectories is reduced by
registration.
For both applications, we use the mean of the starting points of the
trajectories as the reference point $c$ in Definition \ref{defn1}.\vspace*{9pt}

\begin{figure}

\includegraphics{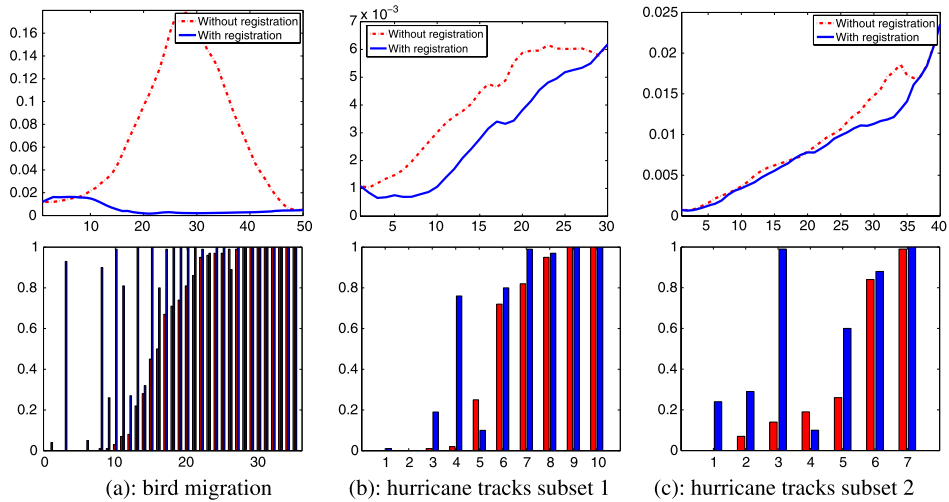}

\caption{$\hat{\rho}$ (1st row) and $p$-values (2nd row) for each trajectory without (red)
and with (blue) registration.} \label{KarcherVariance}
\end{figure}

\textit{Bird migration data}: This data set has 35 migration
trajectories of Swainson's Hawk,
observed during the period 1995 to 1997, each having geographic coordinates measured
at some random times.
Several sample paths are shown at the top row in Figure~\ref{birdmigration}(a).
In the bottom panel of Figure~\ref{birdmigration}(a), we show the
optimal warping functions
$\{\gamma_i^*\}$ used in aligning them and this clearly highlights a
significant temporal variation present in the data.
In Figure~\ref{birdmigration}(b) and (c), we show the Karcher mean
$\mu
$ and the cross-sectional variance $\hat{\rho}$
without and with registration, respectively. In the top row, $\mu$ is
displayed using colors,
where red areas correspond to higher $\hat{\rho}$ value.
In the bottom row, the principal modes of variation are displayed by
ellipses on tangent spaces.
We use the first and second principal tangential directions as the
major and minor axes of ellipses,
and the corresponding singular values as their sizes. We observe that (1)
the mean after registration better preserves the shapes of
trajectories, and
(2) the variance ellipses before registration have their major axes
along the trajectory while
the ellipses after registration exhibit a smaller, actual variability in the data.
Most of the variability after registration is limited to the top end
where the original trajectories indeed have differences.
The top row of Figure~\ref{KarcherVariance}(a) shows a decrease in the
function $\hat{\rho}$ due to the registration.

%

Next we construct a ``Gaussian-type'' model for these trajectories using
estimated summaries for two cases (with and without
temporal registration), as described previously, and compute $p$-values
of individual trajectories using Monte
Carlo simulation.
The results are shown in the bottom of Figure~\ref{KarcherVariance}(a),
where we note a general increase in the $p$-values for the
original trajectories after the alignment. This is attributed to a
reduced variance in the model due to
temporal alignment and the resulting movement of individual samples
closer to the mean values.\medskip

\begin{figure}

\includegraphics{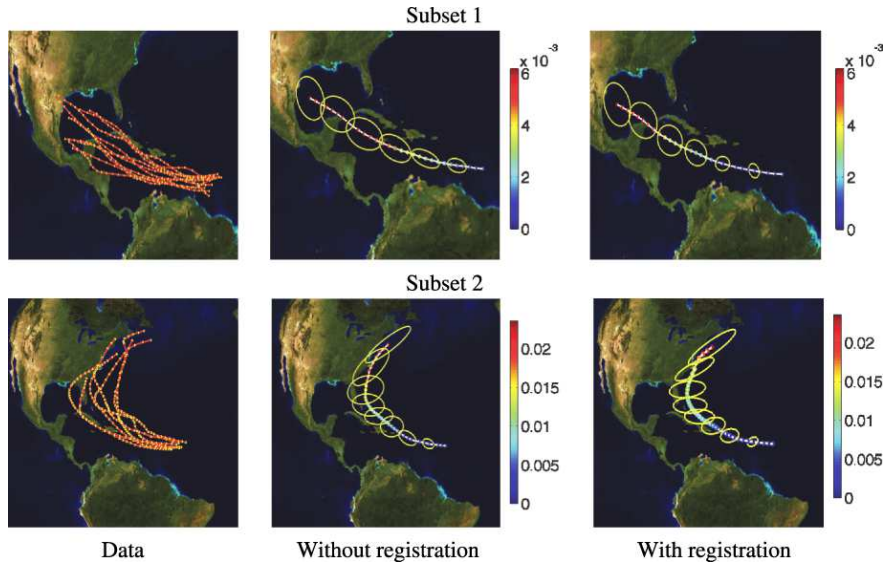}

\caption{Summary of hurricane tracks without and with temporal registration.} \label{hurricanetrack}
\end{figure}

\textit{Hurricane tracks}: We choose two subsets of Atlantic
Tracks File 1851-2011,
available on the National Hurricane Center website.\footnote{\url{http://www.nhc.noaa.gov/pastall.shtml}.}
The first subset has 10 tracks and another has 7 tracks, with
observations at six-hour separation. In Figure~\ref{hurricanetrack} we
show the data, their Karcher mean and
variance without and with registration for each subset. The decrease in the
value of $\hat{\rho}$ is shown in the top of Figure~\ref{KarcherVariance}(b)~and~(c).
Although the decrease here is not as large as the previous \mbox{example}, we
observe about 20\% reduction in $\hat{\rho}$
on average due to registration. In the bottom plots of Figure~\ref{KarcherVariance}(b)~and~(c),
it is also seen that there is a general increase of the \mbox{$p$-}values
after registration, although
they decreased in a few cases. This is because those trajectories
are closer to the mean without registration.
%
%

\section{Vehicle trajectories on $\mathrm{SE}(2)$}\label{sec4}

Here we study the problem of classifying vehicle trajectories into
broad motion patterns using data obtained from traffic videos. While
the general motion of a vehicle at a traffic intersection is
predicable---left turn, right turn, $U$~turn or straight line---the
travel speeds of vehicles may be different in distinct instances
due to traffic variations. Since we are interested in tracking position
and orientation of a vehicle,
we consider individual tracks as parameterized trajectories on $\SE(2)$,
which is a
semidirect product of $\mathrm{SO}(2)$ and $\mathbb{R}^2$, that is,
$\SE(2) = \mathrm{SO}(2)
\rtimes\mathbb{R}^2$. For the rotation component $O \in\mathrm
{SO}(2)$ and
tangent vectors $X_1, X_2 \in T_{O}(\mathrm{SO}(2))$, the standard
Riemannian metric is given by $ \langle X_1,X_2  \rangle =
\operatorname{trace}(X_1^T
X_2)$, while we use the Euclidean metric for $\mathbb{R}^2$. We choose the
rotation component of
$c$ as the identity matrix and the translation component as $[0, 0]$.
We found that the
results of registration, clustering and classification are quite stable
with respect to
different choices of $c$.
For a tangent vector $W \in T_{O}(\mathrm{SO}(2))$,
the parallel transport of $W$ from $O$ to $I_{2 \times2}$ is $O^TW$. The
formulae for the $\mathbb{R}^2$ component are standard.\medskip

%
\begin{figure}[t]
\includegraphics{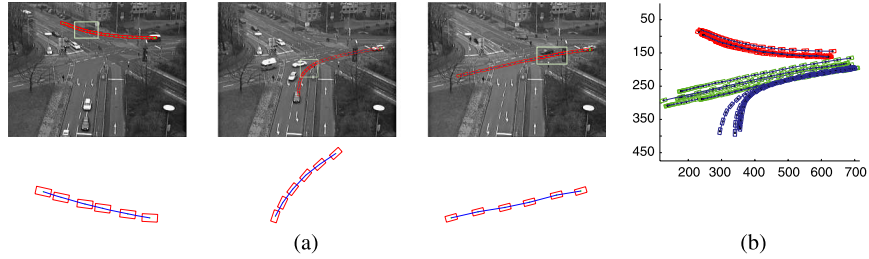}
\caption{\textup{(a)}: Real trajectories in $\SE(2)$ obtained from a traffic video.
\textup{(b)}: Trajectories used for clustering.} \label{figSE2pic}
\end{figure}
%
\begin{figure}[b]
\includegraphics{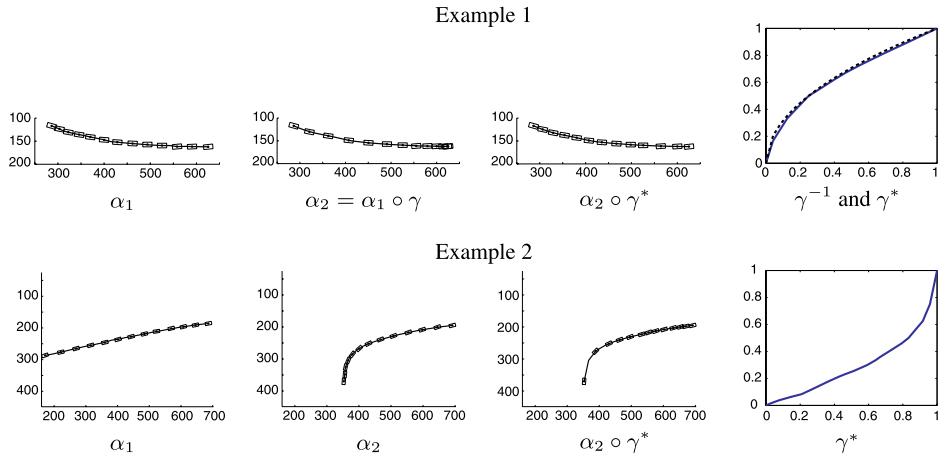}
\caption{Registration of trajectories on $\SE(2)$.}\label{fig2SERegistration}
\end{figure}

\textit{Registration of trajectories}: The data for this
experiment comes from traffic videos available at the Image
Sequence Server website.\footnote{\url{http://i21www.ira.uka.de/image\_sequences/}.}
In Figure~\ref{figSE2pic}(a) we show an example trajectory for each of
the three classes: right turn (first panel),
left turn (second panel), and straight line (third panel).
In this small experiment, the total data includes 14 trajectories with
5 trajectories corresponding to right turn indexed from 1 to 5,
5 trajectories of straight line indexed from 6 to 10, and 4
trajectories of left turn indexed from 11 to 14.

Next, in Figure~\ref{fig2SERegistration} we show two examples of
temporally aligning trajectories described above.
In Example~1 we first choose a trajectory as $\alpha_1$, apply to it a
simulated $\gamma$ and
consider this time-warped trajectory as $\alpha_2$. The right plot of
$\gamma^{-1}$ (dashed) and $\gamma^*$
shows that we are able to recover the simulated time-warping using the
proposed framework.
In Example~2 we show alignment results for trajectories coming from
different classes.
In this case the distance $d_h$ between the trajectories is large,
since they are from different classes,
but it decreases from 14.2 to 10.8 after registration. Furthermore, the
registration result is quite intuitive
since it matches as much of the common features (straight line part) as
possible.\vspace*{9pt}

\begin{figure}[t]

\includegraphics{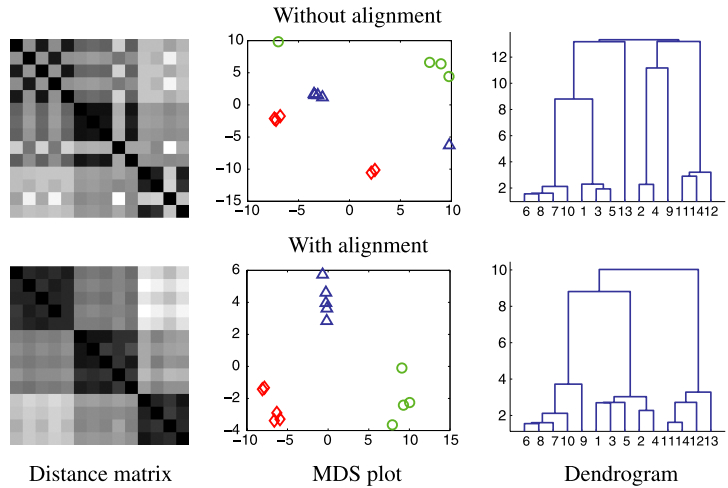}

\caption{Clustering without and with alignment.} \label{cluster}
\end{figure}

\textit{Clustering and classification}:
Here we study the effects of temporal alignment on clustering and
classification.
In the first example, we introduce simple speed
variations in the vehicle motions; these variations represent either
fast-slow or slow-fast movements of a vehicle and apply
them randomly to the 14 given trajectories, shown in Figure~\ref{figSE2pic}(b). In Figure~\ref{cluster} we display the resulting
pairwise distance matrices,
multidimensional scaling (MDS) plots and dendrograms
computed with and without temporal alignment.
The temporal alignment helps in revealing the underlying patterns of
the data.
Also, it greatly improves the clustering performance.

%
%
%
\begin{table}[b]
\tabcolsep=0pt
\tablewidth=200pt
\caption{Classification rates without and with alignment}\label{tab1}
\begin{tabular*}{\tablewidth}{@{\extracolsep{\fill}}@{}lccc@{}}
\hline
\textbf{Classification rate} & \multicolumn{1}{c}{\textbf{1-NN}}& \multicolumn{1}{c}{\textbf{3-NN}} & \textbf{5-NN}\\
\hline
Without alignment & \phantom{0}64.3\%&\phantom{0}64.3\%&50\% \\
With alignment    &100\%\phantom{0.}&100\%\phantom{0.}&93\% \\
\hline
\end{tabular*}
\end{table}

In the second experiment, we introduce more drastic, random speed
variations, corresponding to multiple
stop-and-go patterns of a vehicle. We again apply them to the given
trajectories and compute the distance matrices with and without
temporal alignment.
In Table~\ref{tab1} we report the classification performances based on
1-,~3-~and 5-nearest neighbor (NN) classifiers.
The method described in this paper produces superior classification of
driving patterns. In particular,
we can achieve a~100\% classification rate using the 1-NN classifier.

\section{Shape space of planar contours}\label{sec5}
Motivated by the problem of analyzing human activities using video
data, we
are interested in alignment, comparison and averaging of trajectories on
the shape space of planar, closed curves. There are several mathematical
representations available for this analysis, and we use the representation
of \citet{srivastavaetalPAMI11}. The benefits of using this
representation over other methods are discussed there.
We provide a very brief description and refer the reader to the
original paper for details.
Let $\beta\dvtx \mathbb{S}^1 \mapsto\mathbb{R}^2$ denote a planar
closed curve.
Its corresponding $q$-function is defined as
\[
q(s)= \frac{\dot{\beta}(s) }{\sqrt{ | \dot{\beta}(s)|} },\qquad s \in \mathbb{S}^1.
\]
A major advantage of using $q$-functions to represent shapes of
curves is that the translation variability is automatically
removed ($q$ only depends on $\dot\beta$). To remove the scaling
variability, we re-scale all curves to be of unit length. This
restriction translates to the following condition for
$q$-functions:
$\int_{\mathbb{S}^1}|\dot{\beta}(s)|\,ds=\int_{\mathbb{S}^1}{|q(s)|}^2\,ds=1$.
Therefore, the $q$-functions associated with unit length curves
are elements of a unit hypersphere in the Hilbert space
$\mathbb{L}^2(\mathbb{S}^1,\mathbb{R}^2)$. In order to study shapes
of closed curves,
we impose an additional condition, which ensures that the curve
starts and ends at the same point. This condition is given by $\int_{\mathbb{S}
^1}q(s)|q(s)|\,ds=0$.
Using these two conditions and the $q$-function representation, we
can define the pre-shape space of unit length, closed curves as
\[
{\mathcal C}= \biggl\{q \in\mathbb{L}^2\bigl(\mathbb{S}^1,
\mathbb{R}^2\bigr) \bigg|\int_{\mathbb{S}
^1}{\bigl|q(s)\bigr|}^2\,ds=1,
\int_{\mathbb{S}^1}q(s)\bigl|q(s)\bigr|\,ds=0 \biggr\}.
\]
The shape space of these curves is obtained by removing the
re-parameterization group $\Psi$,
the set of diffeomorphisms from $\mathbb{S}^1$ to itself, and
rotation, that
is, \mbox{${\mathcal S} = {\mathcal C}/(\Psi\times \mathrm{SO}(2))$}.
A unit circle is used as the standard shape and $c$ in Definition~\ref{defn1} is given by its $q$-representation.
For algorithms on computing parallel transports of
tangent vectors along geodesic trajectories in the shape space
${\mathcal
S}$, we refer the reader to
\citet{srivastavaetalPAMI11}.
%
\begin{figure}[t]

\includegraphics{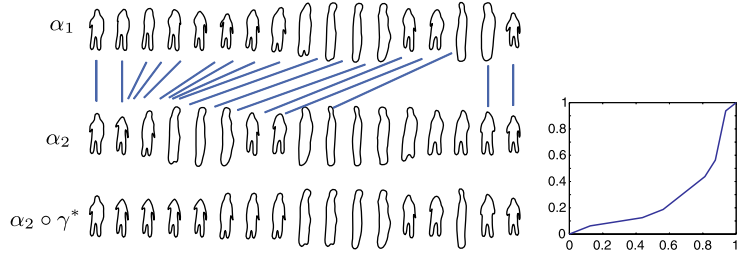}

\caption{Registration of two trajectories on the shape space of planar contours.}\vspace*{-1pt}\label{fig2ShapeRegistration}
\end{figure}
%
%
\begin{figure}[b]

\includegraphics{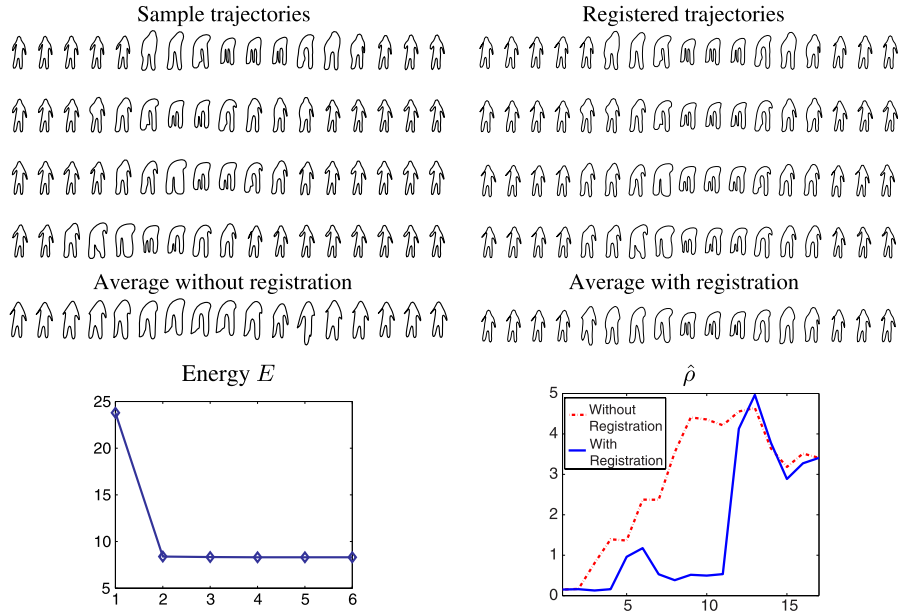}

\caption{Registration and summary of multiple trajectories.} \label{fig2mean1}
\end{figure}

To illustrate our framework, we
apply it to real sequences in the UMD common activities
data set. We use a subset of 8 classes from this data set with 10
instances in each class.
Each instance consists of 80 consecutive planar closed
curves. As a~first step, we down-sample each of these trajectories to
17 contours.\medskip

\textit{Registration}: An example of registering two
trajectories of
planar closed curves from the same class is shown in Figure~\ref{fig2ShapeRegistration}.
The distance $d_h$ between the two trajectories
decreases from 4.27 to 3.26. The optimal $\gamma^*$ for this
registration is shown
in the right panel.\vspace*{9pt}

\textit{Statistical summaries}: We give an example of averaging
and registration of multiple trajectories
using Algorithm \ref{algo1}
in Figure~\ref{fig2mean1}.
The aligned sample trajectories within the same class are much closer
to each
other than before temporal alignment. The energy when computing the
Karcher mean converges quickly,
as shown at the left bottom corner in Figure~\ref{fig2mean1}. The right
bottom plot shows that the cross-sectional variance $\hat\rho$
is significantly reduced after temporal registration.\vspace*{9pt}

\textit{Classification}:
For this activity data set we computed the full pairwise distance
matrix for trajectories, using
$d_h$ (without registration) and $d_s$ (with registration).
The leave-one-out nearest neighbor classification rate (1-NN as
described earlier) for $d_s$ is 95\% as
compared to only 87.5\% when using $d_h$.

\section{Conclusion}\label{sec6}
Statistical analysis of trajectories on nonlinear manifolds is
important in many areas,
including medical imaging and computer vision.
In this paper we have provided a framework for registering, comparing,
summarizing and modeling
trajectories on $\mathbb{S}^2$, $\SE(2)$ and shape space of planar
contours under
invariance to time-warping. Specifically, we have defined a proper
metric, which allows us to register trajectories and compute their sample
means and covariances. For future work, we would like
to extend the framework to other applications with other underlying
manifolds. In addition, we
encourage further efforts on the statistical modeling of such trajectories.



%

\printaddresses

\end{document}